\documentstyle[aps,twocolumn,epsfig]{revtex}

\newcommand \bea{\begin{eqnarray}}
\newcommand \eea{\end{eqnarray}}
\newcommand \ga{\raisebox{-.5ex}{$\stackrel{>}{\sim}$}}
\newcommand \la{\raisebox{-.5ex}{$\stackrel{<}{\sim}$}}
\newcommand{\av}[1]{\langle{#1}\rangle}

\begin{document}
\preprint{\\CERN-TH/2000-368\\LBNL-47191\\hep-ph/0012137} 
\title
{Elliptic Flow at SPS and RHIC: From Kinetic Transport to Hydrodynamics}
\author{P.F.~Kolb$^{a-c}$, P. Huovinen$^{d}$, U.~Heinz$^{a-c}$, 
        and H.~Heiselberg$^e$}
\address{
$^a$Theoretical Physics Division, CERN, CH-1211 Geneva 23, Switzerland\\
$^b$Institut f\"ur Theoretische Physik, Universit\"at Regensburg, 
    D-93040 Regensburg, Germany\\
$^c$Department of Physics, The Ohio State University, 174 West 18th Ave., 
    Columbus, OH 43210, USA\\
$^d$Lawrence Berkeley National Laboratory, Berkeley, CA 94720, USA\\
$^e$NORDITA, Blegdamsvej 17, DK-2100 Copenhagen \O, Denmark}
\date{\today}

\maketitle

\begin{abstract}
Anisotropic transverse flow is studied in Pb+Pb and Au+Au collisions 
at SPS and RHIC energies. The centrality and transverse momentum 
dependence at midrapidity of the elliptic flow coefficient $v_2$ is 
calculated in the hydrodynamic and low density limits. Hydrodynamics 
is found to agree well with the RHIC data for semicentral collisions 
up to transverse momenta of 1--1.5 GeV/$c$, but it considerably 
overestimates the measured elliptic flow at SPS energies. The low 
density limit LDL is inconsistent with the measured magnitude of $v_2$ 
at RHIC energies and with the shape of its $p_t$-dependence at both RHIC 
and SPS energies. The success of the hydrodynamic model points to very 
rapid thermalization in Au+Au collisions at RHIC and provides a serious 
challenge for kinetic approaches based on classical scattering of 
on-shell particles.
\end{abstract}

\smallskip



\medskip

{\it 1. Introduction.--}
Anisotropic flow has been measured in relativistic nuclear collisions
\cite{Bevalac,AGS,NA49v2,NA49QM99,STAR}, and calculations, in particular 
for collisions at SPS and RHIC energies, exist within a variety of 
frameworks: hydrodynamics \cite{Ollitrault,Kolb,KSH00,TS,Hirano}, 
the low density limit of kinetic theory \cite{HL}, parton cascades 
\cite{Zhang,Molnar}, hadronic cascade codes 
\cite{Sorge,RQMD,Xu,Bravina,Soff,Bleicher}, combinations thereof 
\cite{TLS00}, and jet quenching \cite{Wang}. Recent detailed SPS 
\cite{NA49v2,NA49QM99} and RHIC \cite{STAR} data on the centrality and 
transverse momentum dependence of pion and proton elliptic flow begin 
to discriminate between different model calculations and to yield 
quantitative insights into initial conditions, compression, rescattering 
time scales, expansion dynamics and possible phase transitions during the 
expansion stage of the reaction zone.

Cascade calculations based on the incoherent scattering of classical 
on-shell particles and the low density limit of classical kinetic 
theory are expected to work best for peripheral collisions where the 
density of produced particles is sufficiently low and only a few 
rescatterings occur. Central collisions produce higher particle 
densities where the hydrodynamic limit may be more suitable. One of 
the most interesting questions in the kinetic theory of relativistic
heavy ion collisions is where and how the transition between the 
dilute and dense limits happens. 

In this paper we present detailed calculations of the impact parameter
and transverse momentum dependence of elliptic flow in a hydrodynamic
model \cite{Kolb,KSH00} and in the low density limit (LDL) of 
Ref.~\cite{HL}, and we compare the results to recent SPS and RHIC 
data \cite{NA49v2,NA49QM99,STAR}. We test the validity of these two 
models in semi-central Pb+Pb and Au+Au collisions where the elliptic
flow signal is large and thus presents a perfect tool.

{\it 2. Geometry and Anisotropic Flow.--}
Consider two nuclei of radius $R$ colliding at impact parameter $b$.
We refer to the collision as ``central'' when $b\,\la\, {1\over 2} R$, 
``semi-central'' when ${1\over 2}R\,\la\, b\,\la\, {3\over 2}R$, and 
``peripheral'' when $b\,\ga\,{3\over 2}R$. In non-central collisions 
the initial overlap zone is narrower parallel than perpendicular to the 
impact parameter. A simple ansatz for the initial density in the 
interaction region is that it scales with the number of participating 
nucleons per unit area in the transverse plane 
\cite{Ollitrault,Kolb,KSH00,TS}. This predicts a nearly linear dependence 
of the multiplicity on $N_{\rm part}$ in central and semi-central Au+Au 
collisions at RHIC \cite{HKHE}. Other initializations have been proposed
\cite{EKT00}. The sensitivity of elliptic flow to the details of the 
initialization is studied elsewhere \cite{HKHE}. 

\begin{figure}
\vspace*{-1.2cm}
\epsfxsize=8.6truecm
{\centering\mbox{\psfig{file=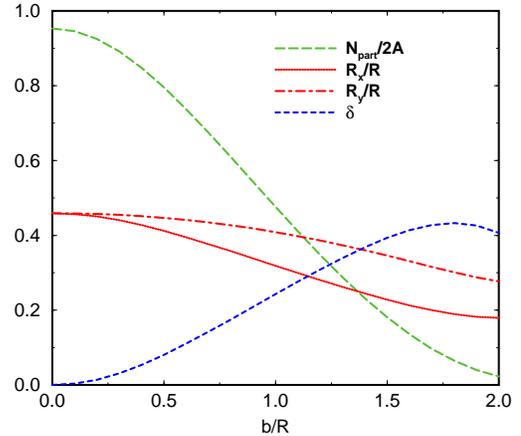,height=8.cm,angle=-90}}}
\caption[]{The number of participants $N_{\rm part}$, the initial 
           transverse radii $R_x, R_y$, and the initial spatial 
           deformation $\epsilon_x\equiv\delta$ for Au+Au at 
           $\sqrt{s}=130\,A$\,GeV, as functions of the scaled 
           impact parameter $b/R$. For details see text.
} 
\label{F1}
\end{figure}

The nuclear thickness function entering the calculation of the initial
conditions \cite{Ollitrault,Kolb,KSH00} is evaluated with realistic 
Woods-Saxon nuclear density profiles for the incoming nuclei, with 
radius $R=1.12\,A^{1/3}-0.86\,A^{-1/3}$\,fm ($A$=207 for Pb+Pb at 
the SPS and $A$=197 for Au+Au at RHIC) and standard surface 
diffuseness $a=0.54$\,fm \cite{Kolb}. The nucleon-nucleon cross 
section is taken as 32\,mb at $\sqrt{s}=17$\,GeV and 40\,mb at 
$\sqrt{s}=130$\,GeV.

With the same parametrization of the initial transverse density 
profile we compute the initial transverse source radii parallel
($R_x(b)= \av{x^2}^{1/2}$) and perpendicular ($R_y(b)= \av{y^2}^{1/2}$)
to the impact parameter. These as well as the initial spatial deformation
\bea
  \epsilon_x(b)\equiv\delta(b) = 
  \frac{R_y^2-R_x^2}{R_y^2+R_x^2}  \label{delta}
\eea 
are shown in Fig.~\ref{F1} for Au+Au at $\sqrt{s}=130\,A$\,GeV as 
functions of the scaled impact parameter $b/R$. 

Particles are initially produced azimuthally symmetric in momentum space. 
The initial deformation in coordinate space generates an azimuthally 
asymmetric momentum distribution if and only if the produced particles 
rescatter off each other. The more rescatterings take place and the 
larger the initial deformation $\delta$, the more anisotropic the final
momentum distribution will be. A quantitative measure of this anisotropy 
is provided by its harmonic coefficients $v_n$ in an event-by-event 
Fourier expansion with respect to the azimuthal angle $\phi$ \cite{VZ96}. 
$v_2$ is called the ``elliptic flow coefficient''.

To relate the initial spatial anisotropy $\delta$ to the measured
momentum anisotropy $v_2$ one has to model the interactions during the 
fireball expansion. The initial average density and source size is 
larger in central than in peripheral collisions. The particle mean 
free paths $\lambda_{\rm mfp}=1/(\sigma\rho)$, where $\sigma$ is the 
scattering cross section and $\rho$ the time-dependent average density 
of scatterers, vary from a few fm or less initially to infinity at 
freeze-out. Since the ratio $\lambda_{\rm mfp}/R_{x,y}$ is smaller  
over a longer time in central than in peripheral collisions, hydrodynamic 
models are more likely to work in central or semicentral collisions,
especially at high energies where the initial particle densities 
are large. The low density limit (LDL), on the other hand, should apply 
to peripheral collisions, especially at low energies with small initial
particle densities.

{\it 3. The Hydrodynamic Limit.--}
The full hydrodynamic treatment of a non-central collision is a tedious 
3+1 dimensional problem \cite{Rischke,Hirano,4dim}. We reduce the 
complexity of the task to 2+1 dimensions by assuming boost-invariant 
longitudinal flow \cite{Ollitrault,Kolb,KSH00,TS}. This assumption 
limits our description to a region around midrapidity which is, 
however, expected to grow as the collision energy increases.

The evolution of a hydrodynamical system is determined by its initial
conditions and equation of state (EOS). We fix the initial conditions 
as in \cite{Kolb,KSH00} by requiring a good fit to the $p_t$-spectra 
of protons and negatively charged particles in central Pb+Pb collisions 
at the SPS \cite{Kolb}. The SPS initial conditions are scaled to the 
RHIC energy of $\sqrt{s}=130\,A$\,GeV by adjusting the initial energy 
density $\epsilon_0$ (keeping the product $T_0\tau_0$ of the initial 
temperature and thermalization time fixed \cite{KSH00}) until the final 
charged particle pseudorapidity density at midrapidity agrees with the 
published measurement by the PHOBOS Collaboration \cite{PHOBOS}. The 
initial baryon density was chosen to give the ratio $\bar{p}/p=0.65$ in 
the final state \cite{STARprel}. Adjusting the initial baryon density 
at fixed initial energy density has no measurable consequences for the 
developing flow pattern, since the pressure is insensitive to $n_b$ when 
the latter is small \cite{KSH00}. 

\begin{center}
\begin{tabular}{|l|c|c|c|c|c|c|}  \hline
              & \multicolumn{3}{c|}{SPS} & \multicolumn{3}{c|}{RHIC} \\ \hline
$T_{\rm f}$ (MeV) $\approx$ 
              &   120   &   120   &   140   &   120   &  140  &  140 \\  
$\epsilon_{\rm f}$ (GeV/fm$^3$)
              &   0.06  &   0.06  &   0.15  &   0.05  &  0.14 & 0.14 \\ \hline
 EOS          &    Q    &    H    &    H    &    Q    &   Q   &   H  \\ 
$\epsilon_0$ (GeV/fm$^3$)
              &   9.0   &    9.0  &   10.0  &   23.0  &  23.0 & 22.3 \\
$n_{b,0}$ (fm$^{-3})$
              &   1.1   &    1.1  &   1.2   &    0.12 &  0.25 & 0.25 \\
$\tau_0$ (fm/c)
              &   0.8   &    0.8  &   0.8   &   0.6   &  0.6  & 0.6  \\ \hline
$T_0$ (MeV)   &   257   &    238  &   242   &   334   &  334  & 270  \\ \hline
$dN_{\rm ch}/dy (b=0) $
              &   390   &    370  &   420   &   670   &  690  & 685 \\ \hline
$dN_{\rm ch}/dy|_{y=0}$ 
              &   355   &    335  &   385   &   615   &  630  & 625 \\
$dN_{\rm ch}/d\eta|_{|\eta|<1}$
              &   310   &    290  &   325   &   545   &  545  & 545  \\ \hline
\end{tabular} \vspace*{2ex}
\end{center}
Table 1. Freeze-out temperatures, equations of state and initial
conditions for central ($b=0$) collisions employed for the hydrodynamical 
calculations shown in this paper. The last two rows show the final
charged particle multiplicity densities in rapidity and pseudorapidity
for the 6\% most central collisions, to facilitate comparison with the
PHOBOS data \cite{PHOBOS}.

\vspace{5mm}

We use two different equations of state to check how the quark-hadron 
phase transition or its absence affects the flow anisotropy. EOS~Q has 
a first order phase transition to QGP at $T_{\rm c}=165$ MeV whereas 
EOS~H contains only hadronic resonances at all densities. At RHIC energies 
EOS~H leads to unrealistic particle densities in the initial state. 
However, this EOS gives an impression how a system without a phase 
transition would behave; this is important when trying to separate 
specific phase transition signatures \cite{Kolb,KSH00} from generic 
hydrodynamical features. Further details of the construction of these 
equations of state can be found in \cite{Sollfrank}.

There is still no final consensus at which temperature the 
hydrodynamical description breaks down at SPS energies 
\cite{freeze-out}, nor is there any a priori reason why this 
temperature should be the same at SPS and RHIC. In our earlier 
studies of Pb+Pb collisions at the SPS \cite{Huovinen} we found 
EOS~Q to favour a lower freeze-out temperature ($T_{\rm f}\approx 
120$\,MeV) than EOS~H ($T_{\rm f}\approx 140$\,MeV). Still, a fit of 
the pion and proton $p_t$ spectra from 158\,$A$\,GeV Pb+Pb collisions 
\cite{NA49spectra} with EOS~H and $T_{\rm f}\approx 120$\,MeV 
looks quite acceptable \cite{Kolb-thesis}. Therefore we do the 
SPS calculations using EOS~Q with freeze-out energy density 
$\epsilon_{\rm f} = 0.06$ GeV/fm$^3$ (corresponding to 
$T_{\rm f}\approx 120$\,MeV), and we probe the effect of the 
freeze-out temperature on elliptic flow by using EOS~H with 
freeze-out energy densities $\epsilon_{\rm f} = 0.06$ as well as 
0.15 GeV/fm$^3$ (the latter corresponds to $T_{\rm f} \approx 
140$\,MeV).

At RHIC the baryon number density at freeze-out is much smaller than 
at the SPS, and the energy densities corresponding to
$T_{\rm f} = 120$ and 140\,MeV are $\epsilon_{\rm f} \approx 0.05$ and 
0.14\,GeV/fm$^3$, respectively. We do the RHIC calculations for 
EOS~Q with both of these freeze-out energy densities and for 
EOS~H with freeze-out density $\epsilon_{\rm f} = 0.14$\,GeV/fm$^3$.
To convert the fluid variables to particle and resonance distributions
we employ the Cooper-Frye freeze-out prescription \cite{CF}. Subsequent
resonance decays are calculated using the decay kinematics described
in \cite{Koch}. The elliptic flow parameters $v_2(p_t)$ and $v_2$ are 
obtained by Fourier expanding the calculated differential and 
$p_t$-integrated momentum distributions. For $T_{\rm f}\approx 140$\,MeV 
decay contributions reduce the elliptic flow of pions by 18--25\% 
depending on the EOS and impact parameter; the corresponding decrease 
at $T_{\rm f}\approx 120$ MeV is 8--15\%.\footnote{In \cite{Hirano} 
resonance decays were found to reduce the pion elliptic flow by 40\%. 
However, in that work only pions with $50 < p_t < 350$ MeV/c were taken 
into account whereas we include either all $p_t$ (SPS) or only cut out 
$p_t<100$\,MeV/$c$ (RHIC). Resonance decays reduce the pion elliptic 
flow especially at low values of $p_t$; using similar $p_t$ cuts as in 
\cite{Hirano} at SPS energies our results are compatible.} 

\begin{figure}
\begin{center}
    \epsfysize 7.0cm \epsfbox{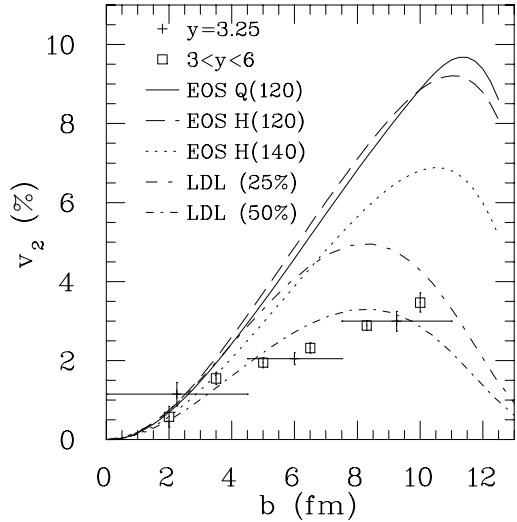}
\caption{Elliptic flow for pions at midrapidity vs.\ centrality, for 
         158\,$A$\,GeV Pb+Pb collisions. Hydrodynamic calculations 
         and results from the LDL are compared to NA49 data 
         \protect\cite{NA49v2,NA49QM99}. Numbers in brackets give 
         $T_{\rm f}$ in MeV and the reduction from resonance decays, 
         respectively.} 
\label{F2}
\end{center}
\end{figure}

\begin{figure}
 \begin{center}
    \epsfysize 7.0cm \epsfbox{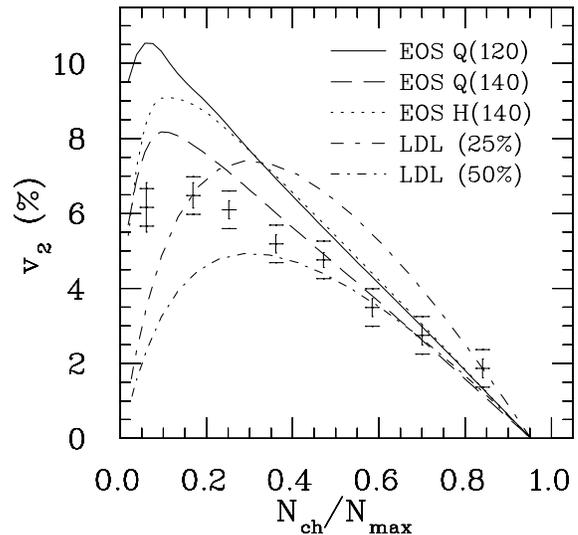}
 \end{center}
\caption{Centrality dependence of the elliptic flow coefficient $v_2$
         for charged particles from Au+Au collisions at $\sqrt{s}=
         130\,A$\,GeV. The data \protect\cite{STAR} are shown with
         the quoted systematic error of $\pm 0.005$. For details see
         text.}
    \label{F3}
\end{figure}

The $p_t$-averaged elliptic flow $v_2$ as a function 
of collision centrality is shown in Fig.~\ref{F2} for Pb+Pb at the SPS 
and in Fig.~\ref{F3} for Au+Au at RHIC. At the SPS energy the 
calculation is for midrapidity pions of all $p_t$ \cite{NA49QM99} 
whereas the RHIC results include all charged particles with 
$|\eta|<1.3$ and $0.1<p_t<2$\,GeV/$c$ \cite{STAR}. We have used the 
same centrality measures (impact parameter $b$ at the SPS, the 
fractional charged particle multiplicity density at midrapidity 
$N_{\rm ch}/N_{\rm max}$ at RHIC) as in the corresponding experimental 
publications.\footnote{To compare with the RHIC data, we had to scale 
$N_{\rm max}$ by a factor 0.95 since the hydrodynamic calculation
cannot account for fluctuations in the charged multiplicity at fixed
impact parameter: while our ``wounded nucleon'' initialization nicely 
describes the measured multiplicity distribution \cite{STAR} up to the 
``knee'' at $b=0$, the calculated distribution drops there abruptly 
to zero while the data show fluctuations up to a value $N_{\rm max}$ 
which is $\approx 5\%$ larger.} 

At the SPS energy the hydrodynamical calculation can reproduce the data 
only for the most central collisions. Already in semi-central collisions 
the calculations overpredict the measured elliptic flow significantly, 
and the disagreement increases to about a factor 2 in peripheral 
collisions. However, the comparison in Fig.~\ref{F2} should be viewed
with some care: as stated above, our calculation applies only to 
midrapidity ($y_{\rm lab}=2.9$) whereas the statistically more 
significant open squares correspond to an average over the forward 
hemisphere. The preliminary data on the rapidity dependence of $v_2$ 
reported in \cite{NA49QM99} show non-trivial structures which are 
averaged out in the open squares. On the other hand, when taking 
into account the $p_t$ cut in \cite{NA49v2}, one finds that the 
midrapidity data from \cite{NA49QM99} at $y=3.25$ (shown as crosses 
in Fig.~\ref{F2}) are significantly lower than the extrapolation of 
$v_2$ to midrapidity published in \cite{NA49v2}. A reliable measurement 
of $v_2$ at midrapidity is thus presently not available for this 
collision system, but would be very welcome.  

Taking the available data at face value, we are unable to account
for them within the hydrodynamic approach even when varying the 
initial conditions, the EOS and the freeze-out temperature within 
the constraints provided by the measured single particle distributions.
Several examples are shown in Fig.~\ref{F2}.

Fig.~\ref{F3} shows that, on the other hand, hydrodynamics successfully
 reproduces the elliptic flow measured at RHIC for central and 
semi-central collisions. The best agreement is reached for EOS~Q
with $T_{\rm f}{\,=\,}140$\,MeV, where discrepancies begin to be significant 
only at impact parameters above 7\,fm, but stay below 20\% even for the 
most peripheral collisions. Lower freeze-out temperatures or the use of 
EOS~H (which is effectively harder at these collision energies, where the
expanding matter spends a large fraction of its total lifetime in the soft
phase transition region \cite{KSH00}) predict somewhat larger elliptic
flow. They also give larger radial flow; when single-particle spectra
from this experiment become available, they will provide a crucial test
of the approach and remove the  ambiguity of the freeze-out temperature 
\cite{HKHVR00}. 

\begin{figure}
\begin{center}
   \epsfysize 7.0cm \epsfbox{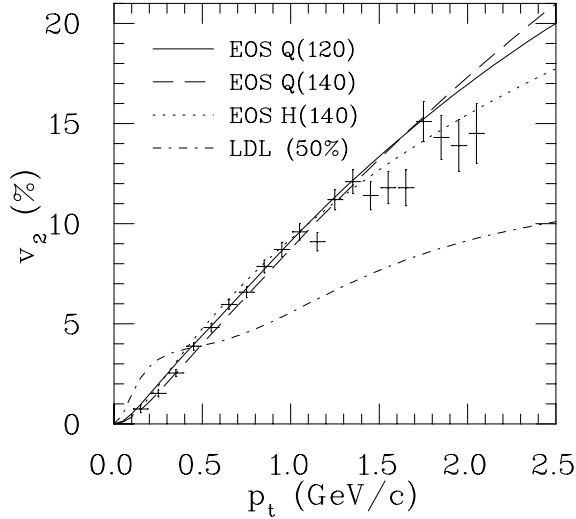}
\caption{The elliptic flow of charged particles from Au+Au collisions 
         at $\sqrt{s}=130\,A$\,GeV vs.\ transverse momentum.
         Hydrodynamic calculations and predictions from the LDL are 
         compared with the data \protect\cite{STAR}. The shape of the 
         LDL curve reflects the weighting of different hadrons with 
         their contribution to the charged particle spectrum; at small
         $p_t$ it is dominated by light pions, at high $p_t$ by heavy 
         baryons.
\label{F4}
}
\end{center}
\end{figure}

Fig.~\ref{F4} shows the $p_t$-dependence of elliptic flow for minimum 
bias Au+Au collisions at RHIC \cite{STAR}. Following the prescription in
\cite{STAR} it is calculated by
 \bea\label{minbias}
   v_2(p_t) = 
   {\int b\,db\,v_2(p_t;b)\, {dN_{\rm ch}\over dy\,p_t\,dp_t}(b)
    \over  
    \int b\,db\,{dN_{\rm ch}\over dy\,p_t\,dp_t}(b)}
 \eea
with a cut-off at $b_{\rm max}=13.5$\,fm. The agreement between the data
and the hydrodynamical calculations is excellent, especially when one 
considers the small variations of the latter upon changing parameters 
within the range allowed by the constraints. Only for $p_t$ above about 
1.5 GeV/$c$ does the measured elliptic flow lag behind the hydrodynamic 
prediction, indicating a departure from thermalization for high-$p_t$ 
particles. In future it will be interesting to follow the data to higher 
$p_t$ where $v_2$ is first expected to saturate due to lack of 
thermalization, before decreasing again as expected from jet quenching 
\cite{Wang}. The hydrodynamic curves start out quadratically at low 
$p_t$, as required by general principles \cite{Dan}, then quickly 
\cite{HKHVR00} turn over to an approximately linear rise and keep 
increasing monotonically with $p_t$, eventually saturating at 
$v_2(p_t)=1$ as $p_t\to\infty$. 

It is not immediately obvious how the good agreement between 
theory and data in Fig.~\ref{F4} is compatible with the clearly 
visible discrepancies in Fig.~\ref{F3} for (semi-)peripheral collisions.
One possibility is that the impact parameter dependence \cite{HKHE} of 
the charged particle multiplicity (i.e. of the normalization of the spectra
which enter the weighting procedure (\ref{minbias})) is different
in theory and experiment. This can be settled by measuring the
dependence of $dN_{\rm ch}/dy$ on the number of participants 
$N_{\rm part}$ \cite{HKHE,EKT00,WG00}, and by providing $v_2(p_t)$
for different centrality bins. A more likely explanation is 
suggested by the observation that the clearly visible differences
in Fig.~\ref{F3} resulting from varying the EOS and $T_{\rm f}$ 
are much smaller in Fig.~\ref{F4}. An analysis shows that the 
variations in Fig.~\ref{F3} result from different slopes of the
{\em single particle spectra\/} which, when averaging the nearly 
identical hydrodynamic curves in Fig.~\ref{F4} over $p_t$, give
different relative weights to the regions of small and large $v_2$.  
The lower $p_t$-averaged $v_2$ for peripheral collisions could thus
be due to steeper single-particle spectra in peripheral collisions
than predicted by the model, e.g. due to earlier freeze-out at 
higher $T_{\rm f}$ and smaller radial flow \cite{TLS00}. This can
be clarified by measuring the single-particle spectra \cite{HKHVR00} 
and the $p_t$-dependence of elliptic flow at different centralities.

\begin{figure}
\begin{center}
  \epsfysize 7.0cm \epsfbox{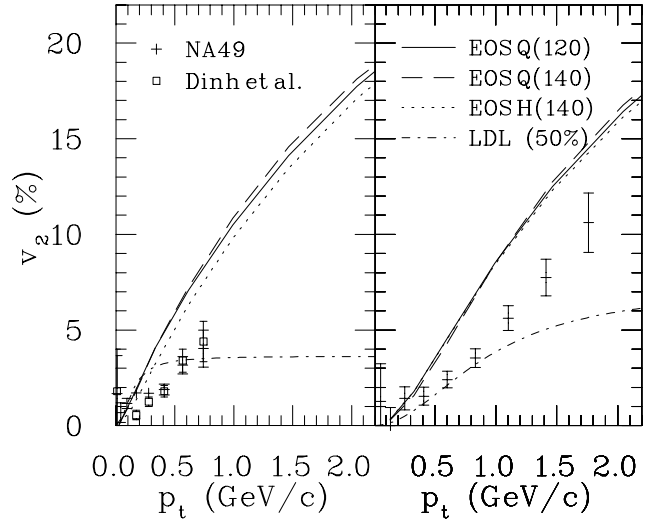}
\caption{$p_t$-dependence of elliptic flow for pions (left) and
         protons (right) from Pb+Pb collisions at $\sqrt{s}=17\,A$\,GeV 
         with impact parameters $b<11$\,fm. The data \protect\cite{NA49v2} 
         correspond to 6.5\,fm\,$<b<$\,8\,fm and are averaged over the 
         forward rapidity interval $4<y<5$ while the hydrodynamic 
         calculations apply to midrapidity $y=2.9$. The squares show
         the NA49 data after correction for azimuthal HBT correlations
         \protect\cite{DBO00}. For details see text.
}
\label{F5}
\end{center}
\end{figure}

The $p_t$-dependence of $v_2$ for pions and protons from semi-central
Pb+Pb collisions at the SPS is shown in Fig.~\ref{F5}. Unfortunately,
no midrapidity data are available, and a meaningful comparison 
between theory and experiment of the magnitude of $v_2$ is thus not 
possible. We note, however, that the data show the same approximately 
linear rise of $v_2$ with $p_t$ and the smaller elliptic flow at small 
$p_t$ for protons than for pions as predicted by hydrodynamics. Such a 
linear rise is inconsistent with the Low Density Limit (LDL) (see below);
it remains to be seen to what extent realistic kinetic codes are able 
to reproduce a linear rise of $v_2(p_t)$, albeit with a smaller slope 
than predicted by hydrodynamics, at SPS energies. 

{\it 4. The Low Density Limit.--}
If the nuclear overlap zone is small and the initial density not
very large, produced particles can escape from the reaction zone 
suffering only a few reinteractions. In the extreme limit the system
is streaming freely, and no collective flow builds up at all. For 
sufficiently dilute systems the elliptic flow can be calculated from
the first order correction to free streaming arising from particle 
collisions. Such a perturbative approach is valid as long as the 
particle mean free paths $\lambda_{\rm mfp}$ are larger than the 
overlap zone $R_{x,y}(b)$. 

In the low density limit (LDL) the effect of scatterings was calculated
to first order by inserting the free streaming distribution into the 
Boltzmann collision term \cite{HL}. The resulting momentum space 
anisotropy leads to an elliptic flow coefficient $v_2^i$ for particle 
species $i$, scattering with particles $j$ ($i,j=\pi,p,K,\dots$),
given by \cite{HL}
 \bea
   v_2^i(b,p_t) = {v_{i\perp}^2(p_t)\over 16\pi}\,
   \frac{\delta(b)\,}{R_x(b) R_y(b)}
   \sum_j {\langle v_{ij}\sigma_{\rm tr}^{ij} \rangle
           \over \langle v_{ij\perp}^2\rangle}         
   \frac{dN_j}{dy}(b) \,.\label{v2c}
\eea
Here, $v_{i\perp}(p_t)$ is the transverse velocity of a particle $i$ 
with momentum $p_t$, and $v_{ij}$ is the relative velocity between 
particle $i$ and scatterer $j$. Brackets $\langle\dots\rangle$ 
denote averaging over scatterer momenta $p_j$. $\sigma_{\rm tr}^{ij}$ 
is the {\it transport cross section} responsible for momentum transfer. 
Similar to the hydrodynamic limit, the elliptic flow coefficient is 
proportional to the initial spatial deformation. Its $p_t$-dependence 
is, however, quite different, saturating at large $p_t$ (when 
$v_{i\perp}\to c$) at values much below 1. At small $p_t$, $v_2$ 
rises quadratically as it should \cite{Dan}, at a rate which is 
directly given by the asymptotic saturation level. This coupling between 
the curvature at small $p_t$ and the saturation level at large $p_t$ 
does not permit a sustained approximately linear rise of $v_2$ in the 
intermediate $p_t$-region.

Let us now try to quantify the parameters entering the LDL formula 
(\ref{v2c}) for SPS and RHIC energies. $\delta(b)$ and $R_{x,y}(b)$ 
are obtained from Fig.~\ref{F1}. We assume that in the region of 
applicability of the LDL the matter can be described with hadronic 
degrees of freedom. For isotropic scattering the transport cross 
sections are about half of the total cross section. However, the 
produced pions are fast and scatter off other pions mostly in the 
forward direction; the same is true for p-wave $\rho$- and 
$\Delta$-resonance scattering. We estimate that the angle-averaged 
cross sections are only $\simeq \langle \sin^2\theta\cos\theta\rangle
= 1/4$ of the total ones, averaged over the relevant relative momentum 
range. For these latter we assume $\sigma^{\pi{\rm M}}\simeq 
10$\,mb, $\sigma^{\pi{\rm B}}=\sigma^{p{\rm M}}\simeq 30$\,mb, and 
$\sigma^{p{\rm B}} \simeq 40$\,mb where M and B stand for arbitrary 
mesonic and baryonic resonances. 

If the cross sections were the same for all particles we could 
replace the sum over $dN_j/dy$ by ${3\over 2} \times dN_{\rm ch}/dy$. 
We use $dN_{\rm ch}/dy\simeq 410$ in central Pb+Pb collisions at the 
SPS ($\sqrt{s}=17\,A$\,GeV) \cite{dNdySPS} and $dN_{\rm ch}/dy \simeq
 630$ for central Au+Au collisions at RHIC ($\sqrt{s}=130\,A$\,GeV) 
\cite{PHOBOS}. [$dN/dy$ at midrapidity is about 15$\%$ larger than 
$dN/d\eta\vert_{|\eta|<1}$.] Since the baryon and anti-baryon cross 
sections are larger, their rapidity densities are required separately; 
we use $dN^{\rm SPS}_{B+\bar{B}}/dy\simeq 110$ for central Pb+Pb 
collisions at the SPS \cite{dNdySPS} and estimate roughly 
$dN^{\rm RHIC}_{B+\bar{B}}/dy\simeq 70$ for central Au+Au collisions 
at RHIC. For the centrality dependence we assume that the rapidity 
densities scale with the number of participants as shown in Fig.~\ref{F1}.  

All relative velocities are assumed to be of the order of the speed 
of light. To obtain the $p_t$-averaged elliptic flow we average over
$v_{i\perp}$ with an exponential $m_t$-distribution with inverse slope 
of 130 MeV; this gives $\langle v_\perp^2 \rangle=0.68$ for pions and
$\langle v_\perp^2 \rangle=0.22$ for protons. In the RHIC data only 
charged particles with $p_t>100$\,MeV/$c$ are included, which we try to 
take into account by increasing $\langle v^2_\perp\rangle$ to 0.75
(assuming pion dominance). 

Finally, we have to correct for resonance decay contributions. The 
actual rapidity density of scatterers in the reaction zone is smaller 
than the observed $dN/dy$ since a large fraction of the latter comes 
from unstable resonances which only decay after the $v_2$-generating
rescatterings have happened. We estimate the corresponding reduction 
factor for the density of scatterers (and thus for $v_2$) to be about 
a factor 2. For illustration of the systematic uncertainties we also 
show curves where only 25$\%$ of the final charged multiplicity arise 
from resonance decays. Clearly, all these numbers are very rough, and 
our estimates for the factor multiplying $v_{i\perp}^2$ in (\ref{v2c}) 
could be easily off by 50$\%$ in both directions.

Curves showing the elliptic flow from the LDL are included in 
Figs.~\ref{F2}--\ref{F5}. The LDL gives about 50$\%$ more elliptic 
flow at RHIC than at the SPS. This reflects mostly the corresponding
ratio of the charged multiplicity densities as shown by the following
expression (see Eq.~(\ref{v2c})):
 \bea
   \frac{v_2^{\rm RHIC}}{v_2^{\rm SPS}} = 
   \frac{dN_{\rm ch}^{\rm RHIC}/dy+c_{\rm RHIC}\,dN_{B+{\bar{B}}}^{\rm RHIC}/dy}
        {dN_{\rm ch}^{\rm SPS}/dy+c_{\rm SPS}\,dN_{B+{\bar{B}}}^{\rm SPS}/dy} 
   \,.
\eea
The constant is given by $c = (2/3) (\sigma^{\pi{\rm B}}_{\rm tr}
/\sigma^{\pi{\rm M}}_{\rm tr}-1)\approx 4/3$. While the LDL thus 
happens to be able to reproduce the impact parameter dependence of the 
$p_t$-averaged elliptic flow at the SPS (Fig.~\ref{F2}), it slightly 
underpredicts the same quantity at RHIC (Fig.~\ref{F3}). Inspection 
of the $p_t$-dependence of $v_2$ in minimum bias events (which is 
dominated by semi-central collisions) reveals, however, that the LDL 
gets it completely wrong at RHIC energies (Fig.~\ref{F4}). Fig.~\ref{F5} 
shows that, while the pion flow data at the SPS are inconclusive due
to their limited $p_t$-coverage, the $p_t$-dependence of $v_2$ for 
protons from semi-central Pb+Pb collisions at the SPS is again 
incompatible with the LDL. Whereas the hydrodynamic model has 
difficulties reproducing the slope (which may, however, be due to 
the different rapidity windows in the data and the model), the
LDL gives a completely wrong shape.   

That the RHIC data for semicentral collisions far exceed the LDL 
prediction demonstrates that first collisions are insufficient and 
multiple collisions are required. The agreement with hydrodynamics
except for a small deficiency for $p_t\ga 1.5$\,GeV/$c$ indicates that
the system is very close to local thermal equilibrium. For very
peripheral collisions we expect that departures from the hydrodynamical 
prediction set in at lower $p_t$-values; it will be interesting to see
whether experiment confirms this.

{\it 5. Conclusions.--}
The recent elliptic flow data from Au+Au collisions at RHIC show 
remarkable quantitative agreement with the hydrodynamical model,
indicating a large degree of thermalization in the earliest collision
stages, well before hadronization. With initial conditions tuned to
data from central collisions at the SPS and no additional adjustment 
of parameters except for a simple scaling of the charged multiplicity 
to the value measured by PHOBOS, the hydrodynamic model reproduces 
quantitatively the centrality dependence of $v_2$ up to impact 
parameters of about 7\,fm and its $p_t$-dependence up to transverse 
momenta of about 1.5 GeV/$c$. Deviations occur only in very peripheral 
collisions and for particles with $p_t>1.5$\,GeV/$c$; they may be due
to a combination of incomplete early thermalization \cite{STAR} and/or
earlier freeze-out \cite{TLS00} in these kinematic regions. The low 
density limit LDL roughly reproduces the shape of the centrality 
dependence of $v_2$ at RHIC, but slightly underpredicts the magnitude 
of the $p_t$-averaged elliptic flow and fails badly for the shape of 
its $p_t$-dependence. It works better for the centrality dependence of 
$v_2$ at the SPS, but again cannot describe the observed nearly linear 
$p_t$-dependence of proton elliptic flow. The hydrodynamic model gets 
the shape of all the $v_2$ distributions at the SPS right, but seems 
to overpredict the absolute magnitude of $v_2$; this last statement is, 
however, uncertain due to the lack of reliable midrapidity data from 
Pb+Pb collisions at the SPS.

These findings suggest that at RHIC thermalization sets in very early 
(the hydrodynamic simulations point to a thermalization time scale of 
less than 1 fm/$c$), but that it may take longer at the SPS. A better
understanding of the onset of deviations from hydrodynamic behaviour
at RHIC, which should be provided by future measurements of 
$dN_{\rm ch}/dy$, $v_2(p_t)$, and the single-particle $p_t$-spectra 
as functions of the number of participants, will yield crucial insights 
into the kinetic evolution at the earliest collision stages. Existing 
parton \cite{Zhang,Molnar} and hadron \cite{Bleicher} cascade 
calculations reproduce the approximate linear rise of $v_2(p_t)$ up 
to $p_t\la 500$\,MeV/$c$, but at higher $p_t$ the elliptic flow levels 
off, and in UrQMD its absolute value at RHIC, averaged over $p_t$, is 
underpredicted by about a factor 4-5 \cite{Bleicher}. The parton 
cascade MPC \cite{Molnar} builds up elliptic flow earlier, but 
quantitatively does not perform very much better. This raises serious 
questions about the adequacy of incoherent scattering among on-shell 
particles to describe the early collision stage and the approach to 
thermalization in ultrarelativistic heavy-ion collisions.

The accurate agreement of the STAR data \cite{STAR} with hydrodynamic 
predictions \cite{KSH00} proves that with elliptic flow one has found 
a hadronic signature which is sensitive to the hot and dense quark-gluon
plasma stage before hadronization sets in. More detailed measurements 
like those mentioned above (in particular the shape of the single-particle 
spectra) should help to confirm and further constrain the picture 
\cite{HKHVR00}. This will open the door to quantitatively characterize 
the QGP equation of state and in particular to distinguish between 
equations of state with and without a phase transition \cite{KSH00}.

We gratefully acknowledge many fruitful discussions with M.~Bleicher, 
V.~Koch, A.~Poskanzer, P.V. Ruuskanen, R.~Snellings, S.~Voloshin, and 
N.~Xu. P.K. wishes to thank both LBNL and BNL for hospitality while finishing
this paper. This work was supported in part by the Director, Office of Science,
Office of High Energy and Nuclear Physics, Division of Nuclear Physics, 
and by the Office of Basic Energy Sciences, Division of Nuclear Sciences, 
of the U.S. Department of Energy under Contract No. DE-AC03-76SF00098.



\begin{references} 

\bibitem{Bevalac}
  W. Reisdorf and H.G. Ritter, Ann. Rev. Nucl. Part. Sci. {\bf 47} (1997) 663.

\bibitem{AGS} 
  J.~Barrette {\it et al.} (E877 Collaboration), Phys. Rev. {\bf C56} 
  (1997) 3254; H. Liu {\it et al.} (E895 Collaboration), Nucl. Phys. 
  {\bf A638} (1998) 451c.

\bibitem{NA49v2} 
  H. Appelsh\"{a}user {\it et al.} (NA49 Collaboration), Phys. Rev. Lett. 
  {\bf 80} (1998) 4136 (in the present work the updated figures archived
  on the NA49 home page 
  http://na49info.cern.ch/na49/Archives/Images/Publica\-tions/ were 
  used)

\bibitem{NA49QM99}
  A. Poskanzer, S. Voloshin {\it et al.} (NA49 Collaboration), 
  Nucl. Phys. {\bf A661} (1999) 341c.

\bibitem{STAR} 
  K.~H.~Ackermann {\it et al.} (STAR Collaboration), nucl-ex/0009011.

\bibitem{Ollitrault} 
  J.Y. Ollitrault, Phys. Rev. D {\bf 46} (1992) 229; {\it ibid.} {\bf 48} 
  (1993) 1132; and Nucl. Phys. {\bf A638} (1998) 195c.

\bibitem{Kolb} 
  P.F.~Kolb, J.~Sollfrank, and U.~Heinz, Phys. Lett. B {\bf 459} (1999) 667;
  P.F.~Kolb, J.~Sollfrank, P.V.~Ruuskanen and U.~Heinz, Nucl. Phys. 
  {\bf A661} (1999) 349c.

\bibitem{KSH00}
  P.F.~Kolb, J.~Sollfrank, and U.~Heinz, Phys. Rev. C {\bf 62} (2000) 
  054909.

\bibitem{TS} 
  D.~Teaney and E.V.~Shuryak, Phys. Rev. Lett. {\bf 83} (1999) 4951.

\bibitem{Hirano}
  T. Hirano, nucl-th/9904082; nucl-th/0004029; 
  T. Hirano, K. Tsuda, and K. Kajimoto, nucl-th/0011087.

\bibitem{HL} 
  H. Heiselberg and A. Levy, Phys. Rev. C {\bf 59} (1999) 2716.

\bibitem{Zhang} 
  B. Zhang, M. Gyulassy, and C.M. Ko, Phys. Lett. B {\bf 455} (1999) 45.

\bibitem{Molnar}
  D. Molnar, talk presented at the 30th International Workshop on 
  Multiparticle Dynamics (ISMD2000), Tihany, Hungary, 9.-15. Oct. 2000, 
  to appear in the proceedings.

\bibitem{Sorge} 
  H. Sorge, Phys. Rev. Lett. {\bf 78} (1997) 2309.

\bibitem{RQMD} 
  R.J.M.~Snellings, A.M.~Poskanzer and S.A.~Voloshin, STAR note 388, 
  nucl-ex/9904003.

\bibitem{Xu} 
  H. Liu, S. Panitkin, and N. Xu, Phys. Rev. {\bf C59} (1999) 348. 

\bibitem{Bravina}
  L.V. Bravina, A. Faessler, C. Fuchs, and E.E. Zabrodin, Phys. Rev. C 
  {\bf 61} (2000) 064902.

\bibitem{Soff} 
  S.~Soff, S.A.~Bass, M.~Bleicher, H.~St\"ocker and W.~Greiner,
  nucl-th/9903061.

\bibitem{Bleicher} 
  M. Bleicher and H. St\"ocker, hep-ph/0006147.

\bibitem{TLS00}
  D. Teaney. J. Lauret, and E. Shuryak, nucl-th/0011058.

\bibitem{Wang}
  X.N. Wang, nucl-th/0009019. 

\bibitem{HKHE}
  P. Huovinen, P. Kolb, U. Heinz, and K.J. Eskola, in preparation.

\bibitem{EKT00}
  K.J. Eskola, K. Kajantie, and K. Tuominen, hep-ph/0009246.

\bibitem{VZ96}
  S.A.~Voloshin and Y.~Zhang, Z. Phys. C {\bf 70} (1996) 665. 

\bibitem{Rischke}
  D.H. Rischke, S. Bernard, and J.A. Maruhn, Nucl. Phys. {\bf A595} 
  (1995) 346; D.H. Rischke, Y. P\"urs\"un, and J.A. Maruhn, {\it ibid.}
  {\bf A595} (1995) 383. 

\bibitem{4dim} 
  C. Nonaka, N. Sasaki, S. Muroya, and O. Miyamura, Nucl. Phys. {\bf A661} 
  (1999) 353c; K. Morita, S. Muroya, H. Nakamura, and C. Nonaka, Phys. 
  Rev. C {\bf 61} (2000) 034904; C.~Nonaka, E.~Honda, and S.~Muroya, 
  hep-ph/0007187.

\bibitem{PHOBOS} 
  B.B. Back {\it et al.} (PHOBOS collaboration), Phys. Rev. Lett. 
  {\bf 85} (2000) 3100.

\bibitem{STARprel}
  Preliminary result reported by the STAR Collaboration at DNP2000,
  see http://www-rnc.lbl.gov/STAR/conf/ talks2000/dnp/ullrich.html

\bibitem{Sollfrank} 
  J.~Sollfrank {\it et al.}, Phys. Rev. C {\bf 55} (1997) 392.

\bibitem{freeze-out}
  B. K\"ampfer {\it et al.}, J. Phys. G {\bf 23} (1997) 2001;
  H.~Appelsh\"auser {\it et al.} (NA49 Collaboration), Eur. Phys. J.
  C {\bf 2} (1998) 661;
  B. Tom\'a\v sik, U.A. Wiedemann, and U. Heinz, nucl-th/9907096;
  T.~Peitzmann, nucl-th/0006025.

\bibitem{Huovinen}
  P.~Huovinen, P.V.~Ruuskanen, and J.~Sollfrank, Nucl. Phys. 
  {\bf A650} (1999) 227.

\bibitem{NA49spectra}
  H. Appelsh\"auser {\it et al.} (NA49 Collaboration), Phys. Rev. Lett. 
  {\bf 82} (1999) 2471.

\bibitem{Kolb-thesis}
  P.F.~Kolb, diploma thesis, University of Regensburg, 1999, unpublished.

\bibitem{CF}
  F.~Cooper and G.~Frye, Phys. Rev. D {\bf 10} (1974) 186.

\bibitem{Koch} 
  J.~Sollfrank, P.~Koch, and U.~Heinz, Z. Phys. C {\bf 52} (1991) 593.

\bibitem{HKHVR00}
  P.~Huovinen, P.~Kolb, U.~Heinz, S.A.~Voloshin, and P.V.~Ruuskanen,
  in preparation.

\bibitem{Dan}
  P. Danielewicz, Phys. Rev. C {\bf 51} (1995) 716.

\bibitem{WG00}
  X.N. Wang and M. Gyulassy, nucl-th/0008014.

\bibitem{DBO00}
  M. Dinh, N. Borghini, and J.-Y. Ollitrault, Phys. Lett. B {\bf 477}
  (2000) 51.
 
\bibitem{dNdySPS}
  P.G. Jones {\it et al.} (NA49 Collaboration), Nucl. Phys. {\bf A610}
  (1996) 188c;
  G. Roland {\it et al.} (NA49 Collaboration), Nucl. Phys. {\bf A638}
  (1998) 91c; 
  F. Sikler {\it et al.} (NA49 Collaboration), Nucl. Phys. {\bf A661}
  (1999) 45c.

\end{references}
\end{document}